\documentstyle{l-aa}

\begin{document}

   \thesaurus{11.07.1; 
              12.07.1} 

   \title{Do the lensing cross-sections of faint galaxies
          cover the whole sky ?}

   \author{Zong-Hong Zhu\inst{1} and Xiang-Ping Wu\inst{2} 
          }

   \offprints{Z. H. Zhu}

   \institute{$^1$Department of Astronomy, Beijing Normal University, 
       Beijing 100875, China\\
     $^2$Beijing Astronomical Observatory, Chinese Academy of Sciences,
       Beijing 100080, China
             }

   \date{Received 00 00, 1997; accepted 00 00, 1997}

   \maketitle

   \begin{abstract}

Very deep galaxy surveys have revealed a considerably large population
of faint galaxies, 
which leads to the speculation that all distant objects
are moderately magnified by the gravitational lensing effects of
galaxies (Fried 1997). In this letter, we present a simple estimate
of the lensing amplitudes by all galaxies up to redshift $z=2$
in terms of galaxy merging and answer the question  
whether the sky is fully covered by the lensing cross-sections 
of galaxies. It is shown that, as a result of the combination of 
the increase of galaxy number density and the decrease of galaxy velocity
dispersion with lookback time,  less than $\sim1/10$ of the
sky to $z=2$ can be moderately affected by galaxies acting as lenses
with magnification $\mu>1.1$. This conclusion is independent of
the galaxy limiting magnitude. In other words, no matter how high the 
surface number density of  faint galaxies becomes, it is unlikely
that their lensing 
cross-sections of $\mu>1.1$ can cover the whole sky.

      \keywords{gravitational lensing --  galaxies: general}

   \end{abstract}

%

\section{Introduction}

It is indeed not a novel idea to suggest that all the distant objects 
may be affected by the gravitational lensing of the matter clumps 
between the sources and  the observer. 
Three decades ago, Barnothy and Barnothy (1968) proposed that
all the quasars were nothing but the gravitationally magnified images of 
Seyfert galactic nuclei. Press and Gunn (1973) showed that the 
probability of the occurrence of gravitational lensing in an $\Omega=1$ 
universe is nearly unity. Unlike the previous speculations
for which there was apparently a lack of both convincing observational
and theoretical supports, 
the current argument is based on the numerous and unprecedented
deep galaxy surveys which have revealed a considerably 
large population of faint galaxies 
(Metcalfe et al. 1996; references therein).  
Using the surface number density of faint galaxies down to $R=26$,
$1.93\times10^{15}/\Box^{\circ}$, Fried (1997) derived the projected 
mean distance between galaxies to be $8^{\prime\prime}.2$, which led
him to the conclusion that all the high redshift ($z>1$) objects are 
moderately magnified by a factor of 1.1--1.5 due to gravitational 
lensing by the intervening galaxies.  Indeed, this was a 
natural and plausible consequence, provided that 
all the faint galaxies were at $z\approx0.5$ and had a mean 
velocity dispersion $\sigma\approx200$--$300$ km s$^{-1}$. 

Nonetheless, spectroscopic redshifts have not been available for most 
of the faint galaxies to date. Namely, we do not yet known where 
these faint galaxies are. For instance, the faint blue galaxies might be
the star-forming galaxies at moderate redshift of $z\sim0.4$ 
(e.g. Broadhurst et al. 1992) or at 
high redshifts of $z\sim2$ (Metcalfe et al. 1996).
While the dispute regarding the merging rate has existed for several years,
it is generally agreed that galaxy mergers may play an important 
role in the formation and evolution of galaxies. At least, 
the merging model provides a good fit to the faint galaxy number counts.
It is particularly noted that the merging alters significantly the 
redshift and velocity dispersion distributions of galaxies.
What is the optical depth due to gravitational lensing for a distant source 
if the redshift and velocity dispersion
information for the faint galaxies according to 
the prediction of galaxy merging is employed ? Can the sky be fully 
covered by the lensing cross-sections of galaxies if faint galaxies 
are neither peaked at $z\sim0.5$ nor distributed randomly in redshift
space ? We would like to answer these questions by modeling the galaxy
matter distribution as the simplest singular isothermal sphere and
the galaxy evolution as merging. Rix et al. (1994)
and Mao \& Kochanek (1994) have presented a sophisticated treatment of
how galaxy mergers affect the various aspects of statistical lensing.   
Here we focus on the specific issue of the lensing covering
of galaxies over the sky.

\section{Galaxy lenses as a result of merging}

Following the idea of Broadhurst et al.(1992), 
we assume that the galaxy merging only increases galaxy number 
with increasing lookback time, whilst maintaining 
the proportion of different types (E, S0, S) of galaxies, their respective
K-corrections and luminosity function shapes. Under these hypotheses, 
the present-day galaxy luminosity function can be written as
\begin{equation}
\phi_i(L_0,0)dL=\phi^*(L_0/L_i^*)^{-\alpha}\exp(-L_0/L_i^*)d(L_0/L_i^*),
\end{equation}
where $i$ indicates the $i-$th morphological type of galaxies: 
$i$=(E, S0, S). The luminosity function at redshift $z$ in the merging model 
is thus
\begin{eqnarray}
\phi_i(L,z)=\phi_i(L_0,0)f(z),\\
f(z)=\exp\{-Q[(1+z)^{-\beta}-1]/\beta\}.
\end{eqnarray}
Here $f(z)$ is  representative of the time-dependence of evolution of 
the luminosity function,  $Q$ is the merging rate and $\beta$ is the
ratio  of the Hubble time $H_0^{-1}$ to the age of the universe. 
The galaxy luminosity $L$ at $z$ is relevant to both the mering rate and
the history of the star formation rate. 
For a matter dominated flat universe of $\Omega_0=1$, 
$\beta=1.5$, while matching the galaxy number counts gives roughly 
$Q\approx4$. This scenario of galaxy merging can account for both 
the redshift distribution and the number counts of galaxies at
optical and near-infrared wavelengths (Broadhurst et al. 1992). 
If we further model the galactic halo by an isothermal sphere
which is characterized solely by its velocity dispersion $\sigma$, 
$\sigma$ at $z$ will be reduced by a factor of $f(z)^{\nu}$ with
respect to its present-day value $\sigma_0$ since the galaxy mass as a result 
of merging would decrease with lookback time. In particular, 
$\nu$ is close to $1/4$ (Rix et al. 1994).
 
The surface number density of faint galaxies to $R=26$ obtained by 
Fried (1997) from the deep observations of the fields around three 
quasars is $1.93\times10^{5}/\Box^{\circ}$, in good agreement with 
the previous surveys(e.g. Metcalfe et al. 1996). This yields a mean 
alignment distance of $4^{\prime\prime}.1$ between the line-of-sight and
the faint galaxies, i.e. $7.3$ $h^{-1}$ kpc in linear size
if the galaxy is at $z=0.5$, where 
$h$ is the Hubble constant in unit of $100$ km/s/Mpc. 
Indeed, assuming that the faint galaxies seen at $R=26$ have a mean 
velocity dispersion of 200 km/s and are located at $z\sim0.5$, 
we can easily estimate that any background sources at $z\sim1$
will be gravitationally magnified by a factor of $\mu>1.1$. So, 
Fried (1997) argued that it is a purely observational fact that
the distant objects must be lensed by foreground galaxies.

Using the empirical formula between the luminosity $L_0$ and central 
velocity dispersion $\sigma_0$ of local galaxies 
$L_0/L_i^*=(\sigma_0/\sigma_i^*)^{g_i}$ with $\sigma^*_i=(225,206,144)$ km/s
and $g_i=(4,4,2.6)$ for $i=(E,S0,S)$ galaxies (see Fukugita \& Turner 1991),
we can compute from eq.(1) the morphological composition \{$\gamma_i$\}
of galaxies by requiring $\sigma_0=200$ km/s. It turns out that 
$\{E:S0:S\}=(62,37,1)$, i.e.,  the galaxies with $\sigma_0>200$ km/s
following the Schechter luminosity function eq.(1)
are mainly composed of the E/S0 populations. As numerous surveys
have shown that spirals are in the majority in the universe ($\sim70\%$), 
the oversimple assumption of Fried (1997) regarding the velocity 
dispersion ($200 - 300$ km/s) for all the faint galaxies 
has overestimated their contributions to gravitational lensing. 
Furthermore, the velocity dispersion of distant galaxies
becomes smaller relative to that of local galaxies in terms of galaxy merging,
which also leads to a decrease of lensing magnification. 
As a consequence, if the faint galaxies
observed by Fried (1997) are $L^*$ spirals at $z\approx0.5$, the 
lensing magnification of a background source at $z=1$ would reduce to  
$\mu\approx1.03$.

\section{Lensing cross-sections}

We now estimate the total lensing cross-sections of galaxies with redshifts 
ranging from 0 to $z_s$ for the distant sources like quasars. For simplicity,
we still employ the singular isothermal sphere for the galaxy matter 
distribution and the evolutionary scenario of galaxy merging.
The lensing cross-section for magnification 
greater than $\mu$ by a single galaxy at $z_d$ is simply 
$\pi\theta_E^2/(\mu-1)^2$ where 
$\theta_E=4\pi[\sigma(z_d)^2/c^2](D_dD_{ds}/D_s)$
is the Einstein radius with $D_{d}$, $D_{ds}$ and $D_s$ being the angular
diameter distances to the galaxy, to the distant source and from the 
galaxy to the source, respectively. The total lensing cross-section by all the
galaxies is 
\begin{equation}
p(z_s,>\mu)=F\;T(z_s)\;\frac{1}{(\mu-1)^2},
\end{equation}
in which 
\begin{equation}
F\equiv 16\pi^3\frac{\phi^*}{cH_0^3}\displaystyle\sum_i
        \gamma_i{{\sigma_i}^*}^4
        \Gamma(-\alpha+4/g_i+1),
\end{equation}
and 
\begin{equation}
T(z_s)\equiv \int_0^{z_s}(1+z_d)^3
  \left(\frac{\tilde{D}_d\tilde{D}_{ds}}{\tilde{D}_s}\right)^2
  f(z_d)^{1-4\nu}d\tilde{r}_{prop},
\end{equation}
where the symbols with a hat of tilde are the corresponding parameters 
in units of $c/H_0$, and $dr_{prop}$ is the proper distance within $dz$ 
of $z$.  Except for the factor of 
$1/(\mu-1)^2$, eq.(4) identifies the eq.(6) of Rix et al. (1994)
for $\Omega_0=1$, in which they concluded that the total optical
depth to multiple images is quite insensitive to merging.
This can be easily shown in terms of eq.(4) 
by noticing that $\nu\approx1/4$. Taking $\nu=1/4$ and 
the galaxy morphological composition $\{\gamma_i\}=
(12\%,19\%,69\%)$ (Postman \& Geller 1984) and 
utilizing the numerical result of $F$ found by Fukugita \& Turner (1991) and
the result of $T$ found by Turner et al. (1984), we have
\begin{equation}
p(z_s,>\mu)=0.047\times\frac{1}{(\mu-1)^2}\frac{4}{15}
            \frac{[(1+z_s)^{1/2}-1]^3}{(1+z_s)^{3/2}}.
\end{equation}
A straightforward computation yields $p(1,>1.1)=3\%$ and
$p(2,>1.1)=10\%$, i.e., the total lensing cross-sections
of $\mu>1.1$ by all the 
galaxies even to $z_s=2$ cannot cover the whole sky at all !
It is important to note that this conclusion is independent of
the limiting magnitudes of the surveys which may reveal a 
remarkably high surface density of galaxies.
Also, our computation has probably overestimated 
the lensing cross-sections of  galaxies 
in the sense that a biasing factor of $\sqrt{1.5}$
between the velocity dispersion of stars and of dark matter particles 
is employed by Fukugita \& Turner (1991) in obtaining $F$ for 
E/S0 galaxies. If such a correction of velocity biasing is unnecessary
(e.g. Kochanek 1994), the total lensing cross-section of galaxies
$p(z_s,>1.1)$ reduces to  $(2\%,6\%)$ for $z_s=(1,2)$.

\section{Conclusion}

The merging model provides an increasing galaxy number 
and a decreasing galaxy mass  with lookback
time, which can relatively easily account for the observed high
surface number density and the redshift distribution 
of galaxies in the deep surveys (Broadhurst et al. 1992).
At least, it works equally well as other models 
(see, for examples, Yoshii \& Sato 1992; Metcalfe et al. 1996).
In the scenario of galaxy mergers, the gravitational lensing of distant 
sources (e.g. quasars) by galaxies is affected by the following
two factors: (1)There will be more galaxies as lenses as one goes back 
in time; (2)The galaxy masses, and equivalently the galaxy velocity
dispersion, will decrease with lookback time. The first factor will
alter significantly the galaxy redshift distributions and enhance the
lensing amplitude, while the second one reduces the lensing 
cross-sections. A combination of these two factors gives rise to
an optical depth to gravitational lensing that is roughly independent of
the galaxy mergers [eqs.(4)-(6); see also Rix et al. 1994; 
Mao \& Kochanek 1994]. 

As a consequence, despite the fact that 
a considerably high surface number density 
of faint galaxies is detected in the deep surveys, the total lensing
cross-sections of galaxies towards a distant source are still 
rather small, and can never fully cover our sky up to $z=2$. The claim
that all the high redshift ($z>1$) objects are moderately magnified
by galaxies (Fried 1997) arises from the oversimple assumptions about
the galaxy redshifts and velocity dispersions 
($200 - 300$ km s$^{-1}$) at high redshifts. We find that the maximum lensing
covering by galaxies to $z=2$ is only $10\%$, and this number is likely
to reduce to $6\%$ for a more realistic galaxy distribution.  Other more
sophisticated models of galaxy evolution should be employed in order
to give a better estimate of the lensing covering by the faint galaxies 
over the sky.

\begin{acknowledgements}
We gratefully acknowledge an anonymous referee for helpful criticisms.  
This work was supported by the National Science Foundation of China.
\end{acknowledgements}

\end{document}